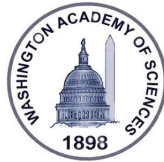

# Hydrogen Clouds from Comets 266/P Christensen and P/2008 Y2 (Gibbs) are Candidates for the Source of the 1977 "WOW" Signal

Antonio Paris
St. Petersburg College

Evan Davies
The Explorers Club, 46 East 70th St, New York, NY

**Published Winter 2015**

### Abstract

On 1977 August 15, the Ohio State University Radio Observatory detected a strong narrowband signal northwest of the globular star cluster M55 in the constellation Sagittarius (Sgr) [1]. The frequency of the signal, which closely matched the hydrogen line (1420.40575177 MHz), peaked at approximately 23:16:01 EDT [2]. Since then, several investigations into the "Wow" signal have ruled out the source as terrestrial in origin or other objects such as satellites, planets and asteroids. From 1977 July 27 to 1977 August 15, comets 266P/Christensen and P/2008 Y2 (Gibbs) were transiting in the neighborhood of the Chi Sagittarii star group. Ephemerides for both comets during this orbital period placed them at the vicinity of the "Wow" signal [3]. Surrounding every active comet, such as 266P/Christensen and P/2008 Y2 (Gibbs), is a large hydrogen cloud with a radius of several million kilometers around their nucleus [4]. These two comets were not detected until *after* 2006, therefore, the comets and/or their hydrogen clouds were not accounted for during the "Wow" signal emission. Because the frequency for the "Wow" signal fell close to the hydrogen line, and the hydrogen clouds of 266P/Christensen and P/2008 Y2 (Gibbs) were in the proximity of the right ascension and declination values of the "Wow" signal, the comet(s) and/or their hydrogen clouds are strong candidates for the source of the 1977 "Wow" signal.

### Introduction

**ON 1977 AUGUST 15** at approximately 23:16:01 EDT, the Big Ear Radio Telescope at The Ohio State University detected an *intermittent* narrowband radio signal (<10 KHz) northwest of the globular star cluster M55 in the constellation of Sagittarius (Sgr) and approximately 2.5° south of the Chi Sagittarii star group [5]. Determining the exact location where the 72-second signal originated from in the sky was problematic because the telescope used two separate feed horns to search for radio signals [5]. The data from the signal, moreover, was processed in such a way that it was difficult to establish which of the two horns detected the signal [2]. There are, therefore, two possible right ascension values for the source of the alleged extraterrestrial intelligence signal: $19^h22^m24.64^s \pm 10^s$ and $19^h25^m17.01^s \pm 10^s$ and the declination was determined to be −27°03′ ± 20 (Table 1) [2]. Two similar values for the signal's frequency were assigned: 1420.356 MHz and 1420.4556 MHz. These two frequencies fall close to the hydrogen line, which is 1420.40575177 MHz [6].

Table 1: Right Ascension and Declination Equinox
Conversions; and Galactic Coordinates for the "Wow" Signal
(Source: Ohio State University Big Horn Report)

|  | Declination | Positive Horn | Negative Horn |
|---|---|---|---|
| **B1950.0 Equinox** | $-27°03' \pm 20'$ | $19^h22^m24.64^s \pm 10^s$ | $19^h25^m17.01^s \pm 10^s$ |
| **J2000.0 Equinox** | $-26°57' \pm 20'$ | $19^h25^m31^s \pm 10^s$ | $19^h28^m22^s \pm 10^s$ |
| **Galactic Latitude** | N/A | $-18^d53.4^m \pm 2.1^m$ | $-19^d28.8^m \pm 2.1^m$ |
| **Galactic Longitude** | N/A | $11^d39.0^m \pm 0.9^m$ | $11^d54.0^m \pm 0.9^m$ |

## Previous Investigations by the Astronomical Community

Subsequent research to re-detect and identify the "Wow" signal by The Ohio State University, the Very Large Array, and The University of Tasmania's Mount Pleasant Radio Observatory were null. After a search of the area where the "Wow" signal was detected (Table 2), the Very Large Array and The Ohio State University Radio Observatory concluded there was strong evidence *against* the origin of the source as terrestrial in nature or objects such as planets, man-made spacecraft, artificial satellites, and radio transmissions emanating from Earth. Furthermore, the Very Large Array proposed the intermittent "Wow" signal matched the signature of a *transiting* celestial source [5], while The University of Tasmania suggested the signal was *moving* with the source of the hydrogen line [7].

Table 2: Right Ascension and Declination Observations Grid by the VLA and Ohio State
(Source: VLA and Ohio State)

|  | Date of Search | RA | DEC |
|---|---|---|---|
| **VLA** | 25 SEP 1995 | $19^h21^m28.1s$ to $19^h25^m48^s$ | $-27°41$ to $-26°18$ |
|  | 07 MAY 1996 | $19^h21^m28.1s$ to $19^h25^m48^s$ | $-27°41$ to $-26°18$ |
| **Ohio State U.** | 05 OCT 1998 | $19^h22^m22^s$ | $-27°03$ |
|  | 09 OCT 1998 | $19^h25^m12^s$ | $-27°03$ |
|  | 9-10 APR 1999 | $19^h25^m12^s$ | $-26°48$ |
|  | 17-18 MAR 1999 | $19^h22^m22^s$ | $-27°18$ |
|  | 20-21 MAR 1999 | $19^h25^m12^s$ | $-27°18$ |
|  | 22-23 MAR 1999 | $19^h22^m22^s$ | $-26°48$ |

## Anatomy of a Comet and Its Hydrogen Cloud

The distinctive parts of a comet include the nucleus, coma, dust tail, ion tail, and a hydrogen cloud. Moderately active comets are surrounded by a widespread cloud of neutral hydrogen atoms. The hydrogen is released from the comet when ultraviolet radiation from the Sun splits water vapor molecules released from the nucleus of the comet into the constituent components oxygen and hydrogen [8]. The size of the hydrogen cloud is determined by the size of the comet and can extend over 100 million km in width, such as the hydrogen cloud of comet Hale Bopp [9]. As a comet approaches the Sun, its hydrogen cloud increases significantly. Since the rate of hydrogen production from the comet's nucleus and coma has been calculated at $5 \times 10^{29}$ atoms of hydrogen every second, the hydrogen cloud is the largest part of the comet [9]. Moreover, due to two closely spaced energy levels in the ground state of the hydrogen atom, the neutral hydrogen cloud enveloping the comet will release photons and emit electromagnetic radiation at a frequency along the hydrogen line (1420.40575177 MHz) [10].

### Comets 266P/Christensen and P/2008 Y2 (Gibbs)

From 1977 July 27 to 1977 August 15, Jupiter-family comets 266P/Christensen and P/2008 Y2 (Gibbs) were transiting in the vicinity of the Chi Sagittarii star group and significantly close to the source of the "Wow" signal (Figure 1) [11]. Of significance to this investigation, the purported source of the "Wow" signal was fixed *between* the right ascension and declination values (Table 3) of comets 266P/Christensen and P/2008 Y2 (Gibbs). On their orbital plane, moreover, 266P/Christensen was 3.8055 AU from Earth and moving at a radial velocity of +13.379 km/s; and P/2008 Y2 (Gibbs) was 4.406 AU from Earth and moving at a radial velocity of +19.641 km/s (Figure 2) [3].

Figure 1: Location of Comets 266P and P/2008 from 1977 July 27 to 1977 August 15. (Source: The Minor Planet Center and NASA JPL Small Body Database) [11].

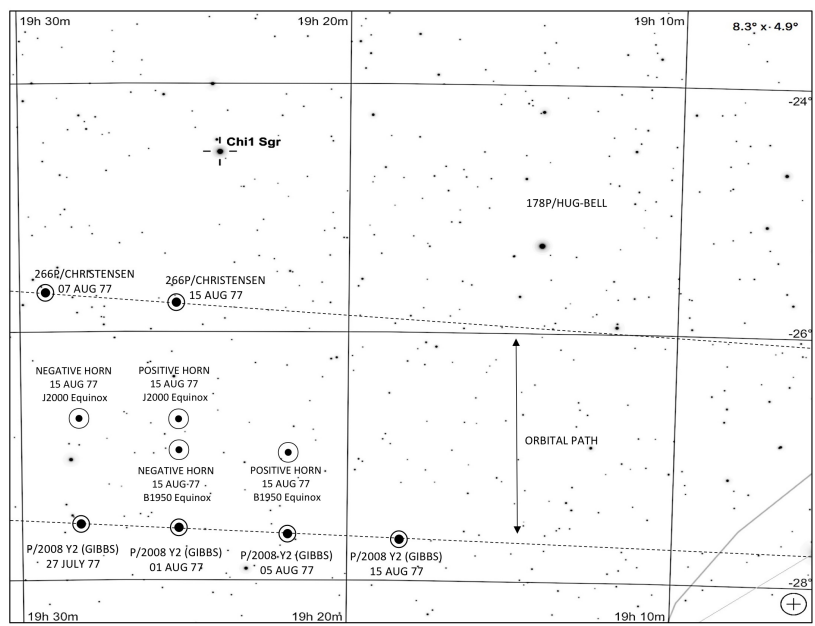

Table 3: Right Ascension and Declination Values for Comets P/2008 Y2 (Gibbs) and 266P/Christensen (Source: Minor Planet Center)

|  | Date | RA | DEC |
|---|---|---|---|
| **P/2008 Y2 (Gibbs)** | 27 JUL 1977 | $19^h28^m12^s \pm 10s$ | -27°31 |
|  | 01 AUG 1977 | $19^h25^m17^s \pm 10s$ | -27°33 |
|  | 05 AUG 1977 | $19^h22^m23^s \pm 10s$ | -27°35 |
|  | 15 AUG 1977 | $19^h16^m37^s \pm 10s$ | -27°36 |
| **266P/Christensen** | 07 AUG 1977 | $19^h29^m47^s \pm 10s$ | -25°53 |
|  | 15 AUG 1977 | $19^h25^m17^s \pm 10s$ | -25°58 |

Figure 2: On 1977 August 15, comet 266P/Christensen was 3.8055 AU from Earth and comet P/2008 Y2 (Gibbs) was 4.406 AU from Earth (Source: JPL Solar System Dynamics Database) [12]

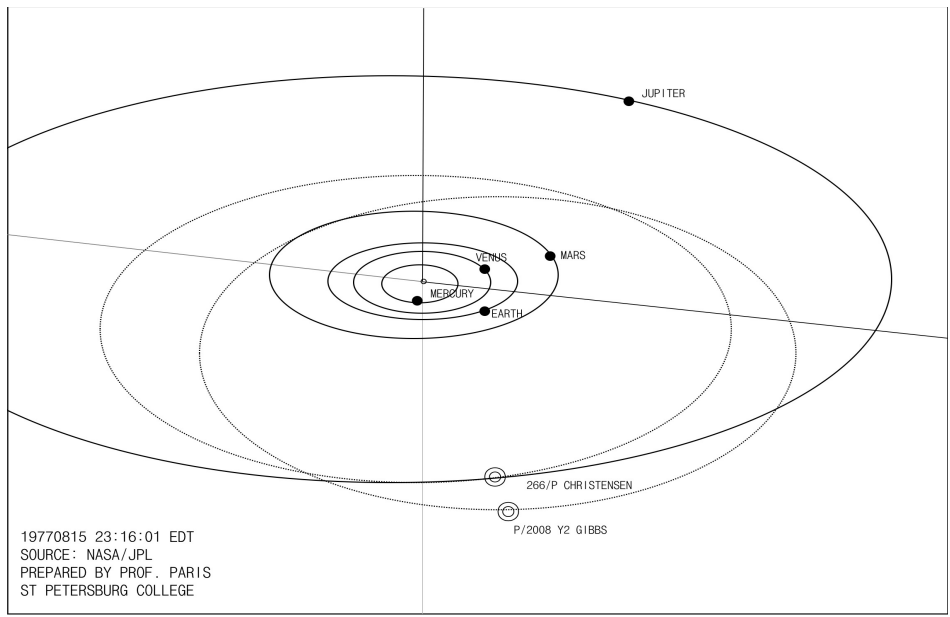

The data regarding comets 266P/Christensen and P/2008 Y2 (Gibbs), therefore, strongly suggest either comet, or both, could be the source of the hydrogen line signal detected by the Ohio State University on 1977 August 15. Chemicals in comets emit radio waves. The hydrogen radio waves from a comet, such as from 266P/Christensen and P/2008 Y2 (Gibbs), travel through space akin to light. Therefore, radio telescopes, including the Big Ear Radio Telescope at The Ohio State University, could have intercepted them. It is noteworthy to comment, moreover, during observations of the area by the Very Large Array and The Ohio State University Radio Observatory (from 1995 to 1999), comet 266P/Christensen and P/2008 Y2 (Gibbs) were *not* in the neighborhood of the right ascension and declination values of the "Wow" signal (Table 4) [5], thus the hydrogen cloud from these two comets would not have been detected. Additionally, because the period for comet 266P/Christensen is 6.63 years and P/2008 Y2 (Gibbs) is 6.8 years [3], their orbital period could account for why the "Wow" signal was intermittent and not detected during subsequent searches of the area.

## Conclusions

There is noteworthy data to propose that the hydrogen signal detected by the Big Ear Radio Telescope at The Ohio State University, specifically 1420.356 MHz and 1420.4556 MHz, emanated from the neutral hydrogen clouds of comets 266P/Christensen and/or P/2008 Y2 (Gibbs). There are, conversely, many unknowns the astronomical community will need to address to confirm the hydrogen clouds from these comets were the source of the 1977 "Wow" signal. To date, no observations have acquired and measured the size, mass and spectral signature, most critically, of these two comets. Additionally, in 1977 the Big Ear Radio Telescope was operating in drift scan mode. Consequently, if a comet (or any celestial object) was the source of the "Wow" signal, it should have been detected in the trailing beam after detection in the leading beam [13]. Comet 266P/Christensen will transit the neighborhood of the "Wow" signal again on 2017 January 25 and can be located at $19^h25^m15.00^s$ and declination $-24°50'$ at a magnitude of +23 [3]. On 2018 January 07, comet P/2008 Y2 (Gibbs) will also transit the neighborhood of the "Wow" signal. Comet P/2008 Y2 (Gibbs) can be located at right ascension $19^h25^m17.6^s$ and declination $-26°05'$ at a magnitude of +26.9 [3]. During this period, the astronomical community will have an opportunity to direct radio telescopes toward this phenomenon, analyze the hydrogen spectra of these two comets, and test the authors' hypothesis.

Table 4: Location of Comets 266P/Christensen and P/2008 Y2 (Gibbs) During VLA and Ohio State Observations (Source: The Minor Planet Center)

|  | Date | RA | DEC |
|---|---|---|---|
| **P/2008 Y2 (Gibbs)** | 25 SEP 1995 (VLA) | $11^h42^m$ | +00°22' |
|  | 07 MAY 1996 (VLA) | $16^h11^m$ | -32°01' |
|  | 05 OCT 1998 (Ohio) | $20^h12^m$ | -22°45' |
|  | 09 OCT 1998 (Ohio) | $20^h15^m$ | -22°41' |
|  | 9-10 APR 1999 (Ohio) | $22^h02^m$ | -13°19' |
|  | 17-18 MAR 1999 (Ohio) | $21^h48^m$ | -14°42' |
|  | 20-21 MAR 1999 (Ohio) | $21^h50^m$ | -14°31' |
|  | 22-23 MAR 1999 (Ohio) | $21^h51^m$ | -14°24' |
| **266P/Christensen** | 25 SEP 1995 (VLA) | $15^h12^m$ | -20°31' |
|  | 07 MAY 1996 (VLA) | $18^h03^m$ | -27°28' |
|  | 05 OCT 1998 (Ohio) | $22^h01^m$ | -14°09' |
|  | 09 OCT 1998 (Ohio) | $22^h00^m$ | -14°12' |
|  | 9-10 APR 1999 (Ohio) | $00^h24^m$ | +02°35' |
|  | 17-18 MAR 1999 (Ohio) | $23^h56^m$ | -00°42' |
|  | 20-21 MAR 1999 (Ohio) | $23^h59^m$ | -00°17' |
|  | 22-23 MAR 1999 (Ohio) | $00^h03^m$ | +00°08' |

**Bios**

**Antonio Paris** is a Professor of Astronomy at St. Petersburg College, FL; the Director of Planetarium and Space Programs at the Museum of Science and Industry in Tampa, FL; and the Chief Scientist at the Center for Planetary Science – a science outreach program promoting astronomy, planetary science, and astrophysics to the next generation of space explorers. He is a member of the Washington Academy of Sciences, the American Astronomical Society, the St. Petersburg Astronomy Club, FL; and the author of two books, *Aerial Phenomena* and *Space Science*.

**Evan Davies** is a fellow of both the Royal Geographical Society, The Explorers Club, and his popular space science writing has appeared in Wiley publications as well as Archaeology and Spaceflight magazines. He is the author of *Emigrating Beyond Earth: Human Adaptation and Space Colonization* and has held a lifelong interest in space exploration.